# Controlled pore formation on mesoporous single crystalline silicon nanowires: threshold and mechanisms

**Stefan Weidemann [1,\*], Maximilian Kockert [1], Dirk Wallacher [2], Manfred Ramsteiner [3], Anna Mogilatenko [4], Klaus Rademann [5] and Saskia F. Fischer [1]**

[1] Group Novel Materials, Humboldt-Universität zu Berlin, 10099 Berlin, Germany;
E-Mails: saskia.fischer@physik.hu-berlin.de

[2] Department for Sample Environment, Helmholtz-Zentrum Berlin, 14109 Berlin, Germany;
E-Mails: dirk.wallacher @helmholtz-berlin.de

[3] Paul-Drude-Institut für Festkörperelektronik 10117 Berlin, Germany;
E-Mails: ramsteiner@pdi-berlin.de

[4] Ferdinand-Braun-Institut, Leibniz-Institut für Höchstfrequenztechnik, 12489 Berlin, Germany;
E-Mails: anna.mogilatenko@fbh-berlin.de

[5] Nanostructured Materials, Humboldt-Universität zu Berlin, 10099 Berlin Germany;
E-Mails: klaus.rademann@chemie.hu-berlin.de

\* Author to whom correspondence should be addressed; E-Mail: weidemann@physik.hu-berlin.de
Tel.: +49-(0)30-2093-7968; Fax: +49-(0)30-2093-7760.

**Abstract:** Silicon nanowires are prepared by the method of the two-step metal-assisted wet chemical etching. We have analyzed the structure of solid, rough and porous nanowire surfaces of boron-doped silicon substrates with resistivities of $\rho > 1000$ $\Omega$cm, $\rho = 14$-$23$ $\Omega$cm, $\rho < 0.01$ $\Omega$cm by scanning electron microscopy and nitrogen gas adsorption. Silicon nanowires prepared from highly-doped silicon reveal mesopores on their surface. However, we found a limit for pore formation. Pores were only formed by etching below a critical $H_2O_2$ concentration ($c_{H2O2} < 0.3$ M). Furthermore, we have determined the pore size distribution in dependence on the etching parameters and characterized the morphology of the pores on the nanowire surface. The pores are in the regime of small mesopores with a mean diameter of 9-13 nm. Crystal and surface structure of individual mesoporous nanowires have been investigated by transmission electron microscopy. The vibrational properties of nanowire ensembles have been investigated by Raman spectroscopy. Heavily boron-doped silicon nanowires are highly porous and the remaining single crystalline silicon nanoscale mesh leads to a redshift and a strong asymmetric line broadening for Raman scattering by optical phonons at 520 cm$^{-1}$. This redshift, $\lambda_{Si\ bulk} = 520$ cm$^{-1}$ $\rightarrow$ $\lambda_{Si\ nanowire} = 512$ cm$^{-1}$, hints to a phonon confinement in mesoporous single crystalline silicon nanowire.

**Keywords:** Silicon nanowire; wet chemical etching; mesoporous surface; gas adsorption; Raman spectroscopy, quantum confinement effects

## 1. Introduction

Nanopatterning can improve bulk properties of crystalline materials [1]. Silicon nanowires attract significant interests in bio-chemical-sensors, catalysis and photocatalysis because of their adjustable electrical properties and their enormous surface-to-volume-ratio [2], [3], [4]. Through their modified electrical behavior silicon nanowires have been proposed for high performance field effect transistors [5], whereas their enhanced optical adsorption is advantageous for photovoltaic applications [6], [7]. In particular silicon nanowires with rough surfaces have strongly decreased thermal conductivities and, hence, have been proposed as promising candidates for thermoelectric devices [1], [8].

Silicon nanostructures can be generated by chemical etching in alkaline and acid solutions [9]. In both cases the wafers orientation, etchant concentration and the different etching rates of silicon and silicon dioxide play significant roles for the nanostructure morphology. Etching along crystallographic axis can be enhanced by applying an external electrical potential [10].

There are various ways for the preparation of silicon nanowires like laser ablation, vapor-liquid-solid growth (VLS) and growth by molecular beam epitaxy (MBE) [11]. However these methods may lead to imperfections like dislocations, grain boundaries, site defects and impurities.

The method of metal-assisted chemical etching has the advantage that it is a low-cost fabrication easily scalable to wafer size [12], [13]. The metal-assisted etching differs in the number of etching steps [14]. In the one-step etching the solution containing the catalytic particles is propelling the nanowire etching, whereas in the two-step etching the amount of catalytic particles is limited in a first etching step and in a second etching solution an oxidizing agent affects the nanowire etching. The limitation of catalytic particles promises a better control in silicon nanowire preparation.

Zhang *et al.* [15] presented a two-step wet chemical etching synthesis of silicon nanowires prepared from *p*-doped and *n*-doped silicon substrates with (100) and (111) orientations. Qu *et al.* [16] and Lin *et al.* [17] investigated how etching parameters, like etching time and concentration of etching solutions, affect the formation of *n*-doped silicon nanowires as well as nanowire porosity. Hochbaum *et al.* [18] and Yuan *et al.* [12] showed for the one-step etching process that with decreasing wafer resistivity the nanowires surface roughness/porosity increases. Although the existence of mesopores on etched silicon nanowires from highly boron-doped silicon is known [14], [16], [17], [18], their shape and the detailed formation mechanism of rough surfaces and nanoporous structures on silicon nanowires remain unclear.

Here, we present a systematic analysis of pore evolution by gas adsorption studies for the two-step etching process of silicon nanowires. We have fabricated silicon nanowire ensembles from undoped Si (resistivity $\rho > 1000$ $\Omega$cm), medium boron-doped Si ($\rho = 14\text{-}23$ $\Omega$cm) and highly boron-doped Si ($\rho < 0.01$ $\Omega$cm) and determined pore diameter distribution, pore volume and sample surface area for highly doped nanowires. The influence of etching time and etchant concentration, illumination and temperature on nanowire formation is discussed, and the pore formation in dependence of etchant concentration and time is analysed. Surface morphology, roughness and crystallinity of individual silicon nanowires are investigated by transmission electron microscopy (TEM). The vibrational properties of nanowire ensembles are investigated by Raman spectroscopy.

## 2. Experimental Section

Silicon nanowires are prepared by the method of metal-assisted chemical etching in a two-step etching approach which has been reported previously [14], [16], [17]. After cleaning the silicon wafer with sonication in acetone (10 min), iso-propanol (10 min) and in boiling acetone (10 min), boiling iso-propanol (10 min) organic residues are removed from the wafer with a fresh prepared, boiling Piranha-solution, a mixture of 3:1 $H_2SO_4$ (97%) an $H_2O_2$ (35%) for 10 min. The generated oxid layer is removed by an HF etching (5% HF for 10 min). The sample is rinsed with deionized water.

In the first etching step with a solution of 4.8 M HF and 0.02 M $AgNO_3$ for 60 s the silver (Ag) nanoparticles are placed on the wafer surface. In this step the silicon wafer is coated by a galvanic replacement of silicon through silver. Silver nanoparticles start to sink into the silicon substrate and begin to form silver dendrites.

After 60 s the sample is rinsed with deionized water again and in a second etching solution consisting of 4.8 M HF and 0.1-0.5 M $H_2O_2$ for an etching time of about 1-3 h the silicon nanowires are formed by the HF etching of the silicon in the following way:

At first [14], oxidation of silicon takes place

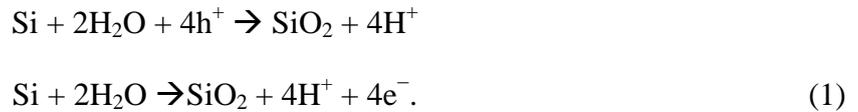

$$Si + 2H_2O + 4h^+ \rightarrow SiO_2 + 4H^+$$

$$Si + 2H_2O \rightarrow SiO_2 + 4H^+ + 4e^-. \tag{1}$$

At second, etching of silicon dioxide produces

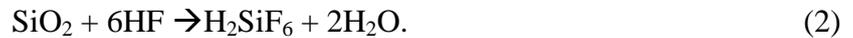

$$SiO_2 + 6HF \rightarrow H_2SiF_6 + 2H_2O. \tag{2}$$

Etching of $SiO_2$ occurs with a higher rate than etching of pure Si in HF. Therefore, it is the dominating etching process [9]. Additionally, the silicon oxidation is promoted by the silver (Ag) and so a higher etching rate of silicon mediated through the catalytic activity of the noble metal is achieved [14], [19].

Through the charge transfer on the interface of silver and silicon the silver nanoparticles are sinking deeper and the silicon nanowires arise as the remainings of unetched silicon. The driving force is the oxidizing power of the oxidizing agent $H_2O_2$.

The driving force is given by

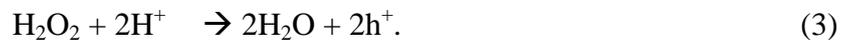

$$H_2O_2 + 2H^+ \rightarrow 2H_2O + 2h^+. \tag{3}$$

$H_2O_2$ oxidizes the Ag particles, which are in turn reduced at the silicon surface by an electron transfer from the silicon and thereby oxidize the silicon. This can be described as a charge transfer process, in which the silver injects holes into the silicon, eq. (1).

After etching the sample is rinsed with deionized water and the Ag particles are removed by a cleaning step in $HNO_3$. The wafer is washed in deionized water for several times and then dried in a nitrogen flow. Normally, all etching steps are performed in the dark and at room temperature.

Here, we prepared nanowires from three types of silicon wafers:

(I) undoped silicon (100) (specific resistivity $\rho > 1000$ $\Omega$cm),

(II) medium boron-doped silicon (100) ($\rho = 14$-23 $\Omega$cm), $p$-type,

(III) highly boron-doped silicon (100) ($\rho < 0.01$ $\Omega$cm), $p$-type.

The as-prepared ensembles of silicon nanowires were characterized by scanning electron microscopy (SEM, Hitachi TM-1000) and details, like the mesoporous surface of single wires, were characterized by SEM (Raith E_ Line plus). Surface area of whole silicon nanowire ensembles was measured by the Brunauer-Emmett-Teller (BET) nitrogen gas adsorption method [20]. The automated gas adsorption station Quantachrome Autosorb-1$^{TM}$ was used. The measurement data were analyzed by Autosorb-1 software to determine the mean pore size distribution (pore diameters and pore volumes) on highly doped nanowires. For this analysis the 1 cm × 1 cm wafers were put into the sample chamber and degassed in vacuum at 140 °C for at least 8 h until the vacuum in the sample chamber has been better than $10^{-4}$ mbar. At T = 77 K nitrogen was dosed in controlled increments into the sample chamber. The pressure equilibrates and the adsorbed nitrogen quantity was calculated. The chamber was filled successively with nitrogen and so the adsorption isotherm was obtained by the adsorbed volume for each relative pressure. Afterwards, the sample chamber was emptied again by successive evaporating enabling to determine the desorption isotherm [21].

Transmission electron microscopy was performed using a JEOL JEM-2200 FS electron microscope. Raman measurements were performed at room temperature in backscattering geometry. The 482.5-nm line of a Coherent Kr$^+$ion laser was focused onto the samples by a confocal microscope. The scattered signal was collected by the same objective and dispersed spectrally by a grating with 600 lines/mm located in an 80-cm Jobin-Yvon monochromator. The signal was recorded using a LN$_2$-cooled CCD with a spectral resolution of 7.5 cm$^{-1}$.

## 3. Results and Discussion

After the first step (HF/AgNO$_3$) of the two-step etching process we observe the formation of small silver particles on the surface of the silicon wafer after a few seconds which is in accordance with [22]. These particles are growing in size with increasing time and some start to form silver dendrites. During the first minutes of etching in the second solution (HF/H$_2$O$_2$) the silver particles sink into the substrate, the silver dendrites grow on and the nanowires are formed as the remainings of the unetched silicon. Longer wires are obtained for increasing H$_2$O$_2$-concentration and longer etching time. This is in accordance with [14], [16], [17]. However, we find that this process is limited and for excessive etchant concentrations and etching times the nanowires are etched on their tips as well.

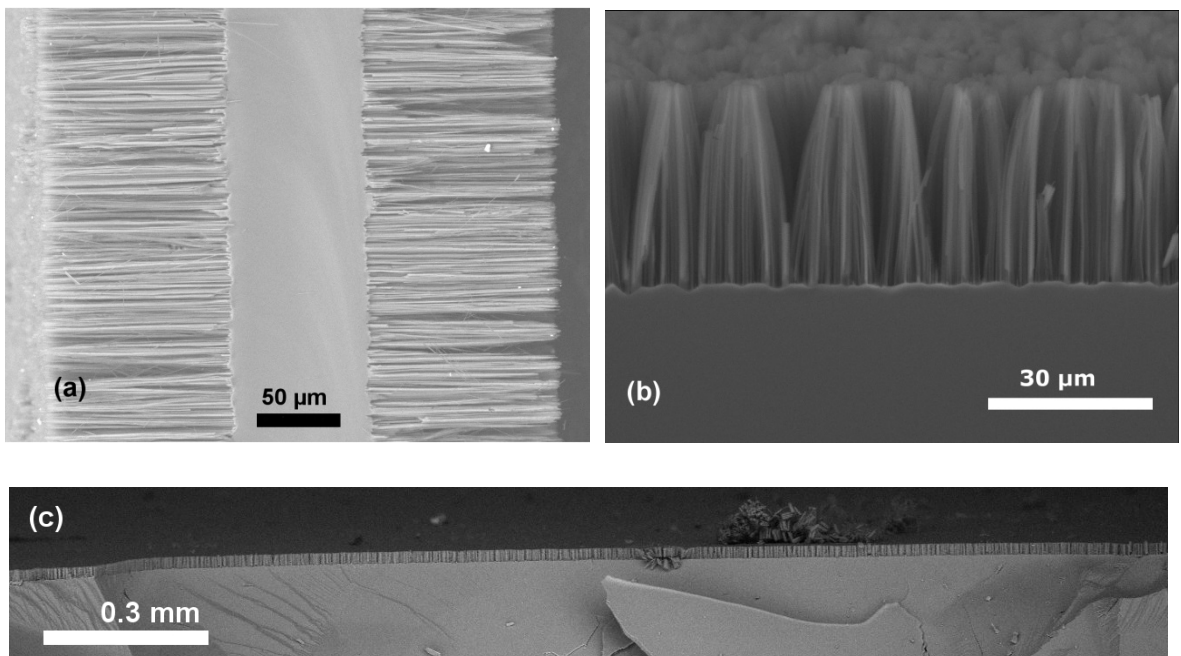

**Figure 1.** Scanning electron micrograph of silicon nanowire ensembles from **(a)** undoped silicon (100) ($\rho > 1000$ $\Omega$cm) c$_{HF}$ = 4.8 M, c$_{H2O2}$ = 0.5 M, etching time 180 min. There are long solid wires (about 110 $\mu$m) on both sides of the thin (< 90 $\mu$m) remaining silicon substrate. **(b)** Highly boron-doped silicon (100) ($\rho < 0.01$ $\Omega$cm) c$_{HF}$ = 4.8 M, c$_{H2O2}$ = 0.3 M, etching time 180 min. The wires

form bundles and lie near against each other. The nanowire tips are bent to the tips of the neighboring nanowires, indicating smaller diameters and pore formation. **(c)** Large scales of uniform silicon nanowire standing on the wafer in cross sectional view.

Figure 1 shows scanning electron micrographs (SEM) of silicon nanowires prepared from different doping concentrations: (a) undoped ($\rho > 1000$ $\Omega$cm) silicon nanowires which are solid, straight and therefore rigid. Approximately 110 $\mu$m long nanowires formed on both sides of the silicon wafer with a remaining width of the wafer of less than 90 $\mu$m are depicted. Morphology of nanowires of the medium boron-doped silicon, not shown, is similar to undoped nanowires. Figure 1(b) shows highly boron-doped silicon nanowires ($\rho < 0.01$ $\Omega$cm). These wires are thinner and more flexible than wires from other silicon types. This is caused by the high length-to-diameter-ratio. The wires form bundles and lean against each other. The nanowire tips are bent to the tips of the neighboring nanowires. This bending could be attributed to a change in the Young's modulus like described before by Lee, *et al.* [23]. Hoffmann, *et al.* [24] and Sohn, *et al.* [25] measured a strongly decreased Young's modulus for silicon nanopillars. Our qualitative bending experiments with an indium tip at single highly-doped nanowires confirm the results in [24] and suggest at least a strong decrease of the Young's modulus. In contrast to [23] and [24] our highly-doped nanowires exhibit a mesoporous surface. A possible influence of the porosity on the Young's modulus demands further investigations, for example by atomic force microscopy bending experiments and force measuring at the nanowires, like in [24]. In figure 1(c) we demonstrate that a homogenous distribution of nanowires by the two-step etching process is feasible on wafer scale.

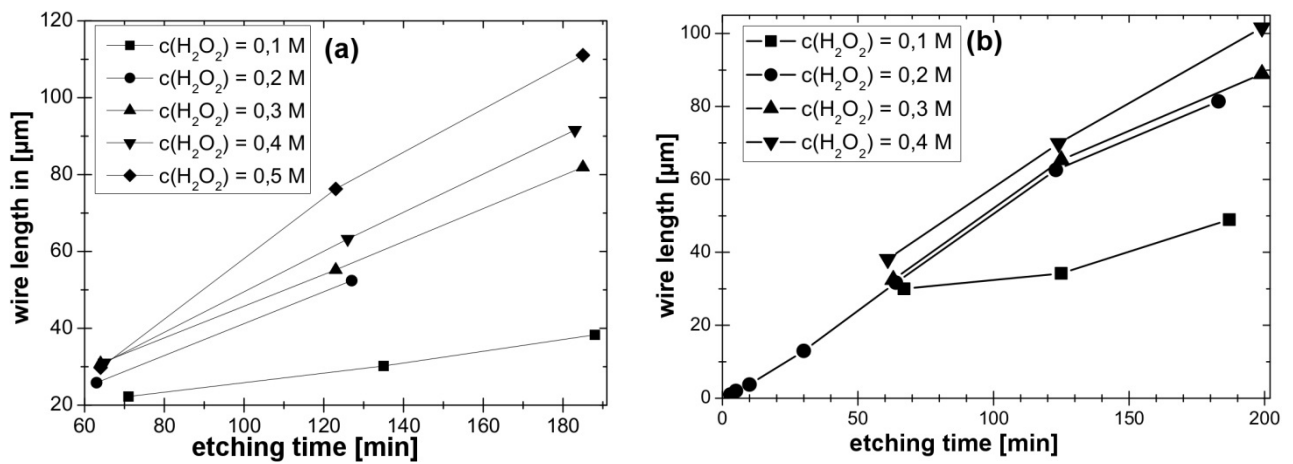

**Figure 2.** Length of the silicon nanowires as a function of the etching time for different concentrations of the oxidizing agent $H_2O_2$. **(a)** Undoped silicon (100), specific resistivity $\rho > 1000$ $\Omega$cm. **(b)** Medium boron-doped silicon (100) ($\rho = 14$-$23$ $\Omega$cm), *p*-type.

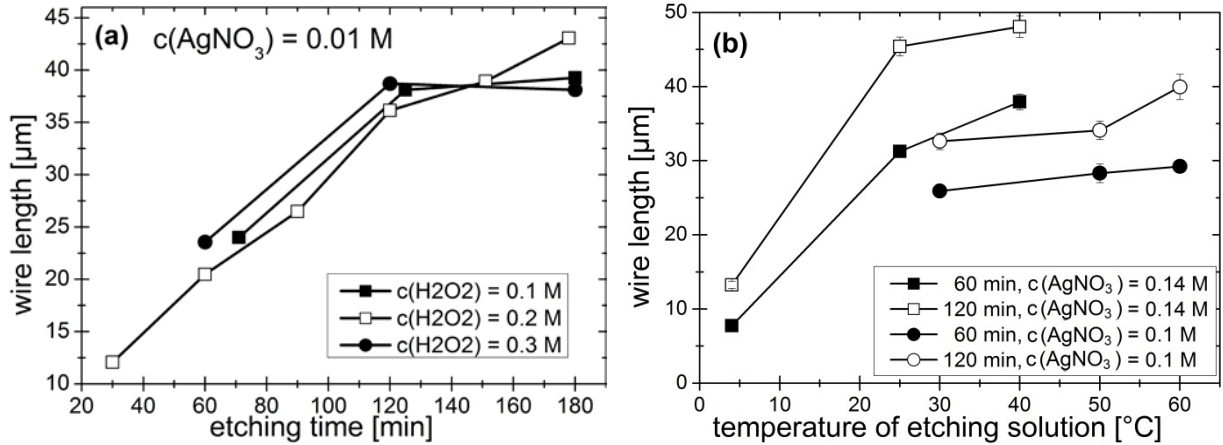

**Figure 3.** Length of the silicon nanowires obtained from highly doped silicon (100) ($\rho$ < 0.01 $\Omega$cm) as a function of the etching time for different concentrations of the oxidizing agent $H_2O_2$ **(a)** and as a function of etching temperatures for different silver nitrate concentration **(b)**. There seems to be a saturation value so that the lengths of the wires are limited to about 50 $\mu$m.

In figures 2 and 3, the main growth trend is presented for nanowires from undoped silicon, fig 2(a), medium boron-doped silicon, fig 2(b), and highly boron-doped silicon, fig 3 (a) and (b). The measurement uncertainty in the length is about 3 - 5 %, which is mainly caused by the inclination and bending of the nanowires. In some cases the nanowires on the bottom side of the wafer are shorter than on the upper side, because the wafers were lying on the bottom of the etching vessel.

In general, we find that longer wires result from higher etchant concentrations and longer etching times, similar to results for $n$-type silicon nanowires [16], [17]. However, high etchant concentrations affect the etching of the nanowire tips: If the tip-etching is as fast or even faster than the sinking of the etching front into the substrate, the nanowire length remains the same or is decreased. This can be seen in the limiting cases of the chart in figure 3(a) where the growth trend is stagnating or even declining, so that the maximal wire length corresponds to about 42 $\mu$m for highly boron-doped silicon. Summing up, nanowire lengths up to 110 $\mu$m (undoped silicon, specific resistivity $\rho$ > 1000 $\Omega$cm), 90 $\mu$m (medium boron-doped silicon, $\rho$ = 14-23 $\Omega$cm) and about 40 $\mu$m (highly boron-doped silicon, $\rho$ < 0.01 $\Omega$cm) are achieved.

Figure 3 (b) shows the nanowire growth trend in dependence of the bath temperature of the etching solution for two different silver nitrate concentrations in the starting solution ($c_{AgNO3}$ = 0.01 M, $c_{AgNO3}$ = 0.014 M) and two etching times (60; 120 min), respectively. It is clear that a higher silver nitrate amount in the first etching solution leads to longer nanowires. Higher etching bath temperatures and illumination during the etching procedure (not shown) also increase the nanowire lengths. These results are in accordance with a recent study on medium boron-doped silicon nanowires ($\rho$ = 10-20 $\Omega$cm) prepared by the one-step etching process [26]. In case of higher $AgNO_3$ concentration we observe higher etching rates during the first etching step which can be explained by the presence of a higher number of catalytic particles promoting the charge transfers at the silicon-solution interface.

Substrate illumination during etching leads to a higher number of photo-excited charge carriers, and a higher etching temperature of the etching solution causes a higher mobility of the etching educts and products, increasing the circulation of the etchant components.

We confirm that the etching occurs along the [100]-axis into the substrate, but also along the other crystallographically equivalent <100> axes, as has been observed before [14], [15], [19], and depicted in fig 4(a). Therefore, we conclude that the Ag particles are not simply sinking into the substrate, but rather that the crystallographic orientation is the dominating factor for the etching direction [14], [19]. This can be explained by the fact that HF-etching of silicon shows a higher etching rate along [100]-crystallographic axis [9], [19].

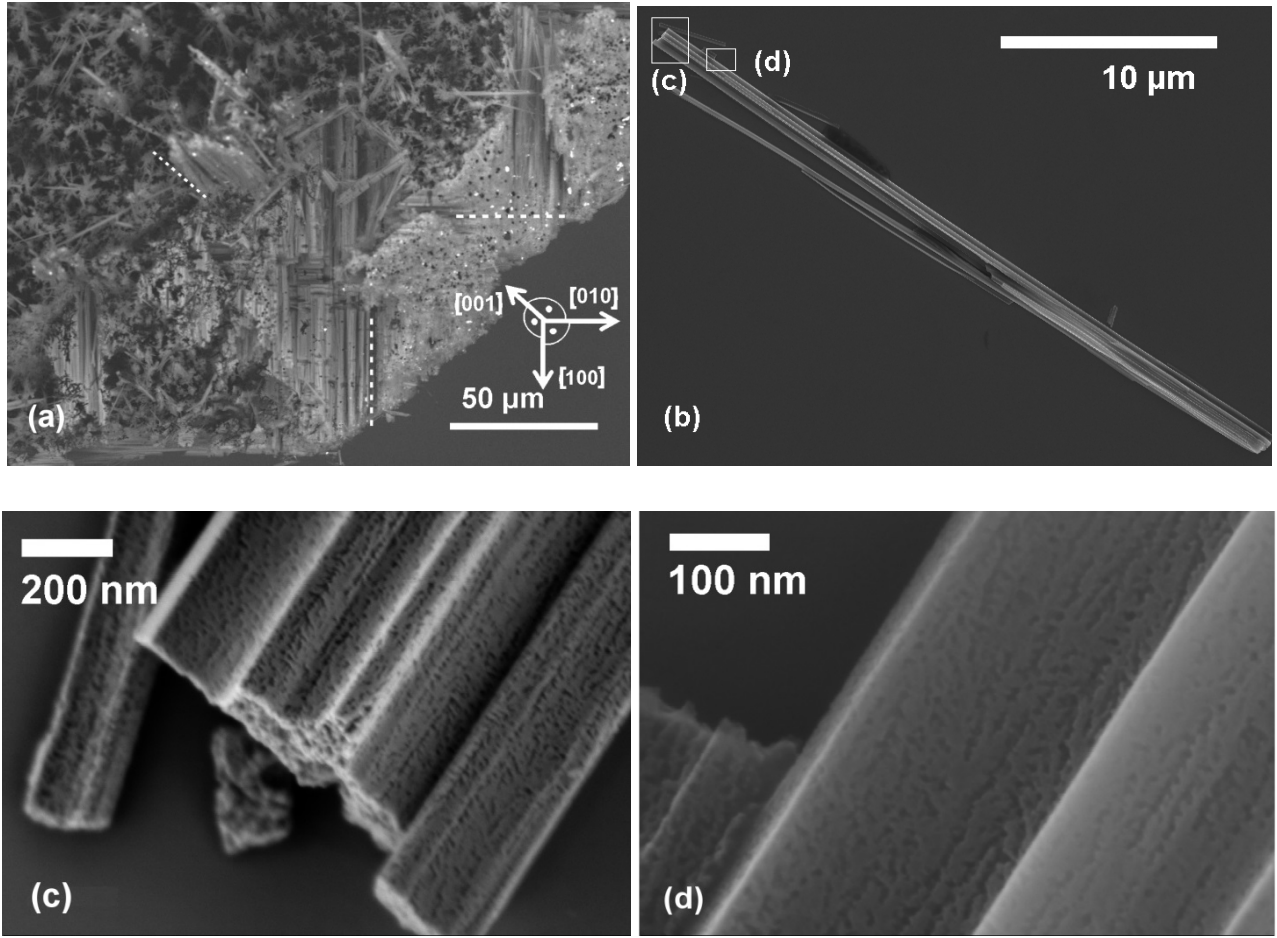

**Figure 4. (a)** Medium boron-doped Si ($\rho$ = 14-23 $\Omega$cm), etched for $t$ = 2 h, $c_{H2O2}$ = 0.2 M, $c_{HF}$ = 4.8 M. SEM image of a slightly tilted silicon wafer after the etching illustrating the etching along the <100>-directions, perpendicular and parallel to the wafer's surface, which has the (100)-orientation. <100>-directions are indicated with white dashed lines. **(b)** SEM image of a single bundle of silicon nanowires from highly boron-doped silicon ($\rho$ < 0.01 $\Omega$cm); marked boxes are magnified in **(c)** showing the cylindrical porous surface structure and **(d)** interconnected pores forming meanderlike trenches on the porous surface.

Figure 4 (a) shows etched structures at the edges of a silicon wafer. The original wafer surface is (100) oriented. Silicon nanowires are standing parallel and perpendicular to the surface. The (100)-directions are indicated with white dashed lines. Figure 4 (b) shows scanning electron micrograph of a bundle of silicon nanowires which are approximated 35 µm long. These nanowires are prepared from highly boron-doped silicon ($\rho$ < 0.01 $\Omega$cm) and have a porous surface structure that can be seen in the magnifications, figure 4 (c), (d). The magnifications of the bottom region of some single wires of this bundle show mesopores of different size and shape. In figure 4 (c) and 4 (d) small mesoscopic (10 nm), cylindrical wholes and dendritic, meanderlike channels are visible on the surface of the nanowires. These patterns can be interpreted as opened (by continued etching) mesopores close to the surface. The total volume and surface of this mesopores, visible and invisible (below the surface) is investigated by gas adsorption measurements as described below.

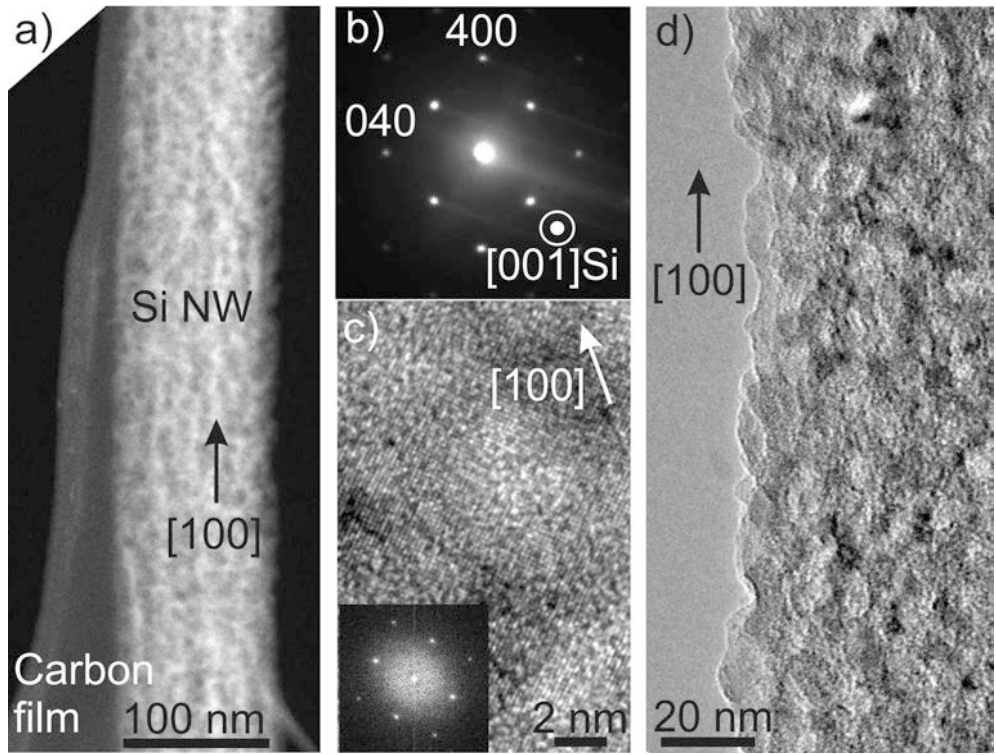

**Figure 5.** Transmission electron microscopy (TEM) images of highly boron-doped, mesoporous silicon nanowires. **(a)** Scanning TEM (STEM) micrograph of a 100 nm thick nanowire. Lying on a carbon film the nanowire exibits a porous surface with a uniform distribution of pores. **(b)** Selected Area Electron Diffraction (SAED) pattern showing the remaining single crystalline structure of the wire. **(c)** High Resolution TEM of the inner part of silicon nanowire with lattice fringes confirming single-crystalline structure; inset: Fourier transform proving the silicon diffraction pattern and the presence of an amorphous amount resulting from the native $SiO_2$ layer around the nanowire. **(d)** TEM image showing the rough and porous surface structure allowing to estimate a pore diameter distribution from 5 to 15 nm and a depth of about 3 nm.

For transmission electron microscopy (TEM) analysis the silicon nanowires were harvested from their substrate, washed in a water solution and dropped on a carbon coated copper TEM-grid. Scanning transmission electron microscopy (STEM) imaging confirms the rough and porous surface structure of the highly doped nanowire along its whole length (figure 5(a)). Selected area electron diffraction (SAED) pattern and high resolution TEM imaging (HRTEM) confirm the remaining single crystalline structure of the wire (figure 5 (b)).

HRTEM reveals that the nanowire core is single crystalline (figure 5(c)). Silicon lattice planes can be seen in the inner part of the wire, weakened in the image by the presence of silicon dioxide and thickness variation ratio around the nanowire. The inset in figure 5(c) shows the Fourier transform confirming the presence of single-crystalline nanowire with some amorphous amount which can be attributed to the $SiO_2$ layer formed at the porous nanowire surface.

TEM-based estimation of the pore diameter and depth, see for example figure 5(d), shows a diameter distribution from 5 to 15 nm (pores of about 8-15 nm in length and 3 nm in depth) in agreement with e.g. Hochbaum *et al.* [18] for one-step etched wires. HRTEM images (not shown) also allowed estimation of the natural silicon dioxide layer thickness ranging from 3 to 6 nm. The surface oxide layer is formed by oxidation of the nanowires in air as well as in aqueous solutions. Dark/bright variations in intensity originate from thickness variation of the wire due to the rough surface.

The porous structure of the nanowire surfaces has been investigated by the method of nitrogen gas adsorption. For this analysis the samples are degassed (at T = 140 °C in vacuum) and the sample chamber is filled with controlled increments of nitrogen starting at a relative pressure of $p/p_0 = 10^{-5}$ at T = 77 K, where $p_0$ is the saturation vapour pressure of liquid $N_2$ at 77 K ($p_0 = 10^5$ Pa). For the adsorption isotherm the adsorbed volume quantity is calculated for each pressure. Decrementing the adsorbed nitrogen amount in the sample chamber reveals the desorption isotherm. The low pressure region of a sorption isotherm corresponds to a mono- and multilayer adsorption regime of the adsorbate on the substrate. From these data the total surface area of the sample can be derived by the method of Brunauer, Emmet and Teller (BET) [21]. By the approaches from Barrett, Joyner and Halenda (BJH) or Density Functional Theory (DFT) we determine the mean pore diameter and the total pore volume. BJH considers that multilayer adsorption could result in capillary condensation under the assumption that the pressure for spontaneous condensation/evaporation of the adsorptive in a cylindrical pore is determined by the pore size according to the Kelvin equation [20], [21]. BJH is recommended for the purpose of comparing the pore sizes among the different materials with the same mesostructures [27]. DFT is modeling interactions and pore geometry by a microscopic treatment of sorption on a molecular level and thereby gives realistic density profiles as a function of temperature and pressure.

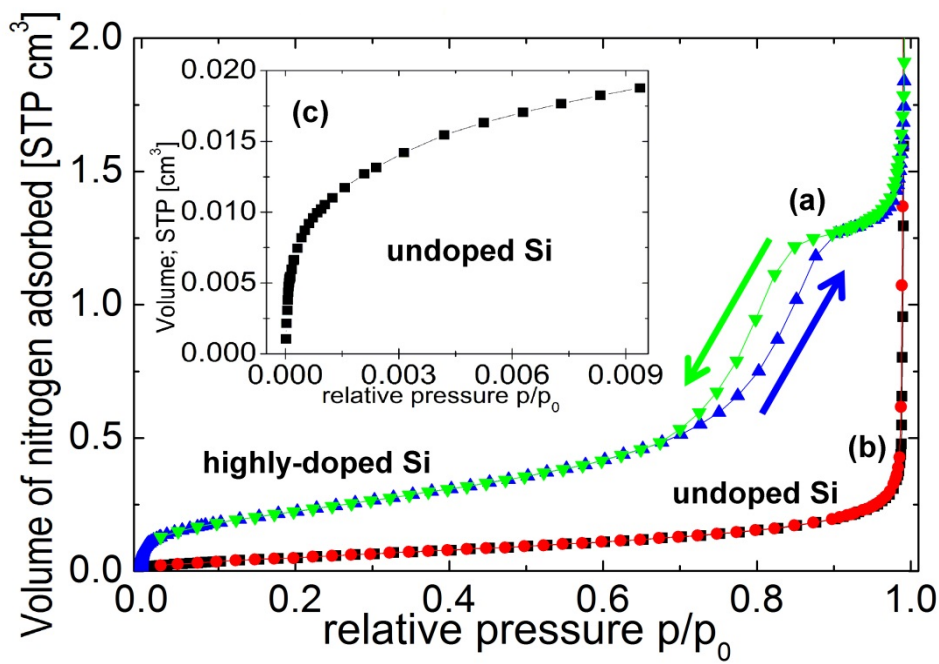

**Figure 6.** Example for nitrogen gas adsorption/desorption isotherms on an ensemble of nanowires of **(a)** highly (blue/green triangles) doped and **(b)** undoped (red/black) silicon. The isotherms of the highly-doped substrate show a hysteresis which is characteristic for mesoporous surface on the nanowires. Silicon nanowires from undoped wafers show no pores. Inset **(c)**: Magnification of the low-pressure region of the adsorption isotherm of the undoped silicon nanowires to illustrate the formation of the first adsorbed nitrogen monolayer, indicated by the knee at $p/p_0 = 0.05$.

Figure 6 depicts two representative gas adsorption isotherms which show a completely different behavior. In the case of undoped silicon, graph (b), the 'knee' (see inset) at $p/p_0 < 0.05$ indicates the formation of a complete monolayer of nitrogen adsorbed on the sample surface. In the relative pressure range of $0.05 < p/p_0 < 0.9$ a further multilayer growth of nitrogen on the surface takes place. The sample's surface area can be calculated by Multipoint-BET-analysis in the pressure range $0.15 < p/p_0 < 0.35$. For higher relative pressures $p/p_0 > 0.9$ the adsorbed gas amount increases very sharply. This rise announces the filling of the space between the nanowires and finally the rest of the sample chamber by liquid nitrogen. This shape corresponds to a type II sorption isotherm which is typical for macro- or nonporous materials, where unrestricted multilayer adsorption can occur [21].

Graph (a) in figure 6 shows the gas adsorption isotherm for silicon nanowires prepared from highly boron-doped silicon wafer. This is a type IV isotherm with its hysteresis loop, which is typical for mesoporous materials [21]. For relative pressures, $p/p_0 < 0.05$ a more pronounced "knee" from the adsorbed nitrogen monolayer is visible. In the range of $0.05 < p/p_0 < 0.6$ multilayer growth appears which will be evaluated by a multipoint-BET fit.

In the regime of $0.6 < p/p_0 < 0.9$ there is a hysteresis loop between the adsorption and desorption isotherm, which indicates capillary condensation of nitrogen in mesopores. From the detailed shape of the isotherm the pore size distribution in the sample can be calculated by the BJH or DFT method. The narrower the pore size distribution is, the steeper is the hysteresis loop corresponding to capillary condensation [21]. The top of the hysteresis marks the amount of nitrogen, where all mesopores are filled, which can be used to calculate the total pore volume of the sample. This hysteresis is similar to a type H1 with a broad pore size distribution, according to IUPAC classification. For relative pressures higher than $p/p_0 > 0.9$ there is a sharp increase like before indicating the filling of the space between the nanowires and of the sample chamber.

| $c(H_2O_2)$ | etching time [min] | Wire length [μm] | BET surface area [m²] | BJH total pore volume [$10^{-3}$ cm] | BJH mean pore diameter [nm] |
|---|---|---|---|---|---|
| 0.1 M | 60 | 26.7 | 0.581 | 1.78 | 8.5 |
| 0.1 M | 122 | 37.3 | 0.701 | 2.22 | 9.9 |
| 0.1 M | 180 | 30.1 | 0.622 | 1.99 | 13.1 |
| 0.2 M | 65 | 18.8 | 0.425 | 1.33 | 10 |
| 0.2 M | 122 | 31.1 | 0.524 | 1.97 | 9.9 |
| 0.2 M | 180 | 38.5 | 1.082 | 2.81 | 8.9 |

**Table 1.** Gas adsorption data for nanowire ensembles prepared from highly boron-doped silicon ($\rho < 0.01$ Ωcm). For comparison undoped nanowire ensemble ($\rho > 1000$ Ωcm, $c(H_2O_2) = 0.5$ M, etching time $t = 185$ min, wire length $= 111.1$ μm) reveals a MBET surface area of 0.113 m².

Nanowire ensemble of medium boron-doped silicon ($\rho$ = 14-23 $\Omega$cm, $c$(H$_2$O$_2$) = 0.4 M, etching time $t$ = 199 min, wire length = 101.6 µm) reveals a MBET surface area of 0.212 m².

Table 1 shows the nitrogen gas adsorption results for silicon nanowire ensembles from highly boron-doped silicon. The surface area (multipoint BET) points out the area of all nanowires and mesopores of the ensemble as well as the area of the substrate itself, which is about 3 cm² and therefore negligible. Although the nanowires from undoped and medium-doped silicon are much longer the total surface area of the ensembles is smaller than for wires of highly-doped silicon. This could be attributed to a higher nanowire density and hence smaller nanowire diameters in the case of highly-doped samples. Also it could be attributed to an additional surface area resulting from the pore surface on the highly doped nanowires. The increasing or stagnating mean pore diameters and their total pore volume are shown for highly-doped silicon nanowire samples prepared with different etching times and H$_2$O$_2$-concentrations. In the case of $c$(H$_2$O$_2$) = 0.1 M the total pore volume and the mean pore diameters are growing from one to two hour etching and the pore volume decrease for longer etching time while the mean pore diameter further increase. In the case of $c$(H$_2$O$_2$) = 0.2 M the total pore volume increases and the mean pore diameters stagnate with etching time. The detailed pore diameter distributions are given in Figure 7.

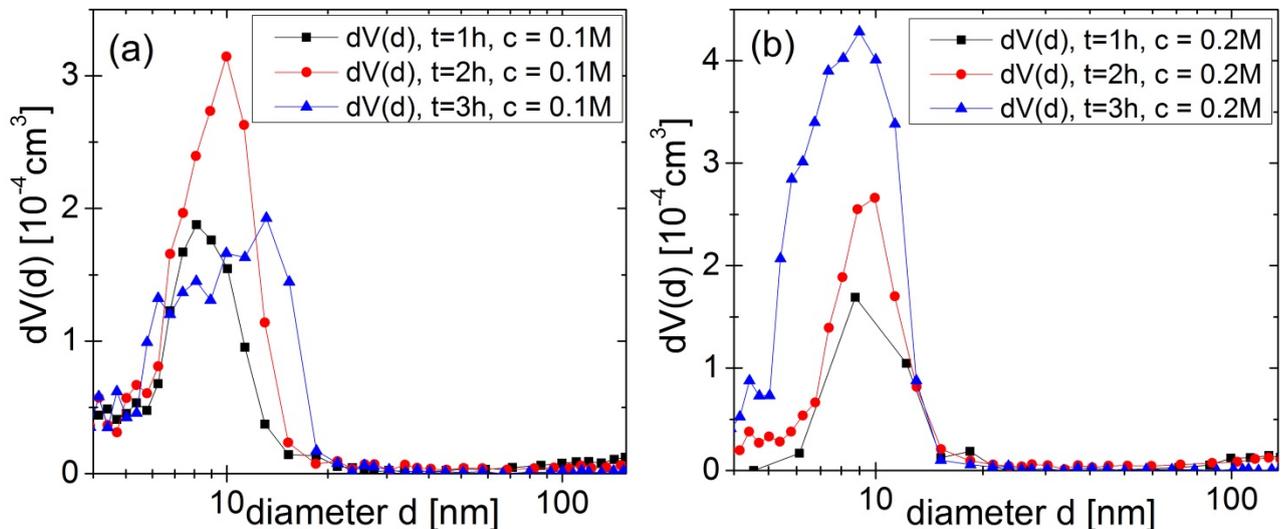

**Figure 7.** Pore size distribution (BJH) for measured isotherms of silicon nanowire ensembles which show hysteresis between ad- and desorption branches. The calculated pore volume fraction $dV$ is plotted versus the pore diameters $d$. The pore size distribution is broadened with longer etching time. Silicon nanowires prepared with c = 0.3 M reveal no pores. **(a)** Highly-doped silicon, preparation parameters: etching time $t$ = 1-3 h and H$_2$O$_2$-concentration $c$ = 0.1 M. **(b)** Highly-doped silicon, preparation parameters: etching time $t$ = 1-3 h and H$_2$O$_2$-concentration $c$ = 0.2 M.

For all measured isotherms of silicon nanowire ensembles which show hysteresis behavior between ad- and desorption branches the pore size distribution has been derived with the BJH-method. DFT-results (not shown) confirm the BJH-results shown in figure 7. The two graphs show the calculated pore volume distribution for related pore diameters for different etching times (1-3 hours) and different H$_2$O$_2$-concentrations, 7(a): $c$ = 0.1 M; 7(b): $c$ = 0.2 M, respectively. As can be seen in fig. 7(a) the pore volume of the sample etched for two hours is increased with respect to the sample etched for one hour and the average pore diameter is shifted to higher diameters. This indicates that existing pores are broadened and deepened and/or that additional pores are generated with bigger diameters. After three hours of etching the peak is much broader and flatter than before. This is a sign for further pore broadening and flattening and the parallel growth of smaller pores. For the higher etching concentration there is a similar situation. After one hour etching there are small mesopores on the ensemble of silicon nanowires. Up to two hours etching time the total pore volume grows and after three hours the pores size distribution is broadened again and the total pore volume is increased further. Silicon nanowires prepared with a $H_2O_2$ concentration $c = 0.3$ M reveal no pores detectable by nitrogen adsorption.

Pore formation seems to originate only near the etching front, because continuous pore forming along the wire would result in a steady increase of the peaks for pores smaller than 9 nm. For longer etching times the pore diameters are broadened, however the pores are not deepened anymore, fig. 7 (a). For $c = 0.2$ M the total pore volume increases continuously and the pore size distribution is broadened for longer etching times. For three hours of etching there are more small pores attributable to slower pore broadening caused by an advanced consumption of hydrogen peroxide. As for the case of the higher concentration ($c = 0.3$ M) the broadening leads to interconnections of the pores leading to a strong surface roughness of the wires. For low etchant concentration the calculations reveal mesopores in the diameter range of 6-18 nm with an average diameter between 9-13 nm.

Our gas adsorption results obtained on two-step prepared silicon nanowires compare well with these published for silicon nanowires prepared by one-step metal assisted etching process [18]. Consistently, we found mesoporous surfaces on highly boron-doped nanowires also for the preparation by the two-step version of the etching and no pores on undoped and additionally on medium-doped silicon nanowires. Hochbaum *et al.* [18] show gas adsorption data for one nanowire sample, which diameter pore range (2-20 nm) and mean pore diameter (9.7 nm) overlaps with our results. The advantage of the two-step etching method is that the amount of silver can be limited during the first etching step , so that it becomes possible to vary another parameter, i.e. the $H_2O_2$ concentration, similar to [17] for porous *n*-type silicon nanowires. Furthermore, we have figured out the pore size distribution depending on etching time and the etchant concentration. With our results we confirm the finding of the scanning and transmission electron microscopy studies of Yuan, *et al.* [12]. There, the porosity increases from the nanowire root (where the wires are connected to the substrate) to the nanowires tip. This is consistent with our gas adsorption results that the pore formation seems to appear only in a certain distance, but near the etching front. Afterwards, the existing pores grow on getting broadened and flattened. The pore flattening can be explained by the thinning of the nanowire with its exposure time in the etching solution. The pore broadening explains the finding in [12], that there are an increasing number of interconnected pores in the middle and upper part of the nanowires.

Our results can help to understand the formation process of porous silicon nanowires. As depicted in [12] silicon nanowires of different doping concentrations are formed by a vertical etching of the silicon substrate promoted by catalytic active silver particles and, like in our case, accelerated by the oxidizing agent hydrogen peroxide. This process is accompanied by a slight thinning of the nanowires depending on their exposure time to the etching solution (nanowire tips are thinner than their roots [12], [14], [17]). For highly boron-doped silicon nanowires this vertical etching is accompanied by a local lateral etching into the nanowire resulting in a porous surface. Its high dopant concentration leads to surface states acting as nucleation sites where the silver ion reduction occurs randomly spread around the nanowire. As shown in [12] for the one-step etching process the pore formation starts some hundreds of nanometers above the etching front. One explanation could be that charge injections (holes, $p^+$) at the etching front locally increase the carrier concentration and holes with a certain mobility in $p$-type silicon move along the already formed nanowire. In some distance near the etching front these injected charges reach the nanowire surface and promote the etching at random points. So formed pores are growing depending on etching concentration and etching time. Our gas adsorption findings confirm our electron microscopy results toward the porous surface structure. Because of the pore shape distribution it is valid to compare gas adsorption calculations for silicon nanowires among themselves, but it is probably doubtful to compare them to other material systems.

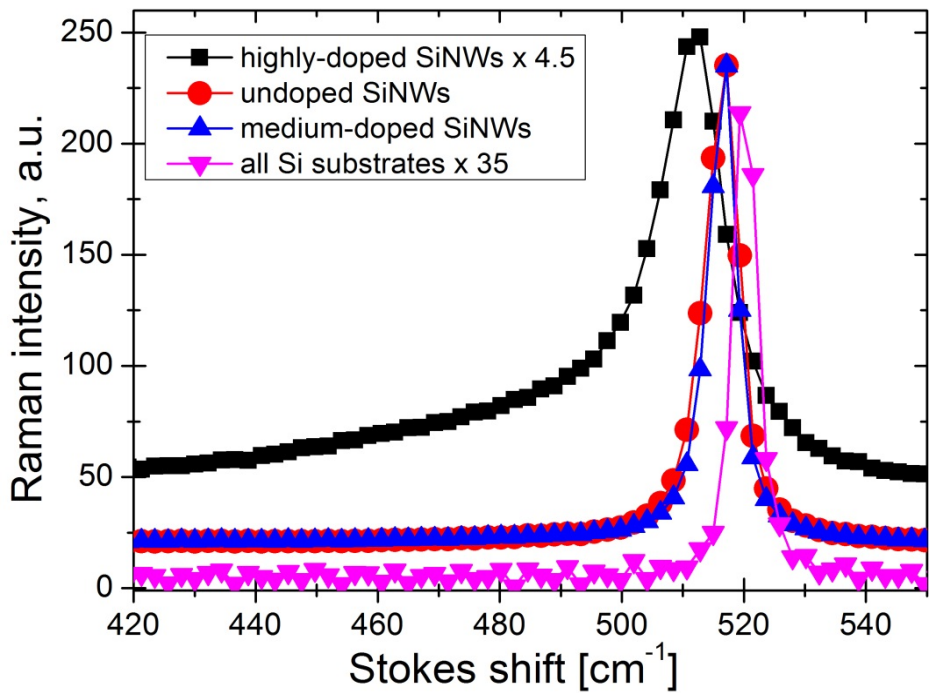

**Figure 8.** Raman spectra of silicon nanowire ensembles prepared from silicon substrates with three different doping levels. The spectrum of a silicon substrate, with the zone-center optical phonon line at 520 cm$^{-1}$, is shown for comparison.

We have investigated the prepared silicon nanowire ensembles with respect to their vibrational properties by Raman spectroscopy. The laser power used for excitation has been reduced to a level for which the influence of local heating is negligible [28]. The Raman spectra shown in figure 8 are dominated by the zone-center optical (O$_\Gamma$) phonon line of silicon at about 520 cm$^{-1}$ (equal to 64 meV) [29], [30]. For undoped as well as medium doped nanowires the O$_\Gamma$ phonon peak exhibits a redshift of 3 cm$^{-1}$ and a moderate broadening as compared to that of the silicon substrates. The modification of the Raman spectrum becomes pronounced for heavily doped nanowires with a redshift of 8 cm$^{-1}$ and a strong asymmetric broadening [full width at half maximum (FWHM) of 15 cm$^{-1}$]. Inhomogeneous strain can be excluded as the origin of the observed spectral changes since the observed redshifts would require the assumption of an unreasonably large magnitude of average strain [31], [32]. In fact, the observed Raman spectra can be explained by the spatial confinement of optical phonons in silicon nanostructures which leads to a relaxation of the pseudo-momentum conservation [28], [33], [34], [35]. Both the observed Raman peak position and FWHM for the heavily doped nanowires can be explained by a model assuming nanospheres with diameters between 3 and 4 nm and a phonon confinement function chosen in analogy to the ground state of an electron in a hard sphere [34]. The phonon confinement can be explained by the formation of a single crystalline silicon nanomesh created by pore formation or a pronounced surface roughness. However, the phonon confinement in the nanowires seems to be not related to the pore formation observed by gas adsorption since nearly identical Raman spectra have been observed for all heavily doped nanowires. Further work is needed in order to clarify this point.

## 4. Conclusions

In our work we report on synthesis and morphology characteristics of silicon nanowires prepared by the two-step electroless etching process. We have prepared nanowire ensembles from different boron- doped substrates and have shown their growth trend and surface morphology which were investigated by scanning electron microscopy and nitrogen gas adsorption; both revealing a mesoporous surface structure on highly doped silicon nanowires. Transmission electron microscopy proves that the structure of mesoporous nanowires remains single crystalline. However, there is a formation of an oxidized surface layer. Mesopores are formed near the etching front and are growing and flattened in dependence of the etching time and etchant concentration. This allows a controlled formation of porous silicon nanowires. Consequences of the nanopatterning to phonon energies and vibrational properties of the nanowires are a redshifted and asymmetric Stokes signal in the Raman spectroscopy for the highly-doped nanowires. The contribution of confinement effects is considered to play the dominating role for this redshift. Effects of the porous surface of the silicon nanowires on their mechanical properties, like the Young's modulus, and their electrical and thermal transport properties demand further investigations.

## Acknowledgments

We gratefully acknowledge Dr. Sven S. Buchholz, formerly at Humboldt Universität zu Berlin, and Raith GmbH for access to E_line_Plus and SEM imaging. S. W. wants to thank Jürgen Sölle and Ulrike Heiden for technical support.

## Conflicts of Interest

The authors declare no conflict of interest.

## References and Notes

1. Boukai, A. I.; Bunimovich, Y.; Tahir-Kheli, J.; Yu, J.-K.; William A. Goddard III, W. A.; Heath, J. R. Silicon nanowires as efficient thermoelectric materials, *Nature*, **2008**, *451* (10), 168- 171; doi:10.1038/nature06458.
2. Oh, J.; Deutsch, T. G.; Yuan, H.-C.; Branz, H. M. Nanoporous black silicon photocathode for $H_2$ production by photoelectrochemical water splitting. *Energy Environ Sci.* **2011**, *4*, 1690-1694.
3. Peng, K.-Q.; Wang, X.; Lee, S.-T. Gas sensing properties of single crystalline porous silicon nanowires, *Appl. Phys. Lett.*, **2010**, *95*, 243112; doi:10.1063/1.3275794.
4. Zhou, X.T.; Hu, J.Q.; Li, C.P.; Ma, D.D.D.; Lee, C.S.; Lee, S.T. Silicon nanowires as chemical sensors, *Chemical Physics Letters* , **2003**, *369*, 220-224; doi:10.1016/S0009-2614(02)02008-0.

5.  Cui, Y.; Zhong, Z.; Wang, D.; Wang, W. U.; Lieber, C.M. High Performance Silicon Nanowire Field Effect Transistors, *Nano Lett.*, **2003**, *3* (2), 149-152; doi:10.1021/nl025875l.

6.  Jia, G.; Höger, I.; Gawlik, A.; Dellith, J.; Bailey, L. R.; Ulyashin, A.; Falk, F. Wet chemically prepared silicon nanowire array on low-cost substrates for photovoltaic applications, *Phys. Status Solidi A*, **2013**, *210* (4), 728-731; doi:10.1002/pssa.201200531.

7.  Sivakov, V.; Andrä, G.; Gawlik, A; Berger, A.; Plentz, J; Falk, F.; Christiansen, S. H. Silicon nanowire-based solar cells on glass: synthesis, optical properties, and cell parameters, *Nano Lett.*, **2009**, *9*(4):1549-54; doi: 10.1021/nl803641f.

8.  Hochbaum, A. I.; Chen, R; Delgado, R. D.; Liang, W.; Garnett, E.C.; Najarian, M.; Majumdar, A.; Yang, P. Enhanced thermoelectric performance of rough silicon nanowires, *Nature*, **2008**, *451* (10), 163-168; doi:10.1038/nature06381.

9.  Lehmann, V. *Electrochemistry of Silicon: Instrumentation, Science, Materials and Applications*, 3rd ed.; Wiley-VCH Verlag GmbH: Weinheim, Germany, 2002.

10. Lehmann, V.; Rönnebeck, S. The Physics of Macropore Formation in Low-Doped *p*-Type Silicon. *J. Electrochem. Soc.* **1999**, *146* , 8, 2968-2975.

11. Schmidt, V.; Wittemann, J.V.; Senz,S.; Gösele, U. Silicon Nanowires: A Review on Aspects of their Growth and their Electrical Properties, *Adv. Mater.*, **2009**, 21, 2681–2702.

12. Yuan, G.; Mitdank, R.; Mogilatenko, A.; Fischer, S. F. Porous Nanostructures and Thermoelectric Power Measurement of Electro-Less Etched Black Silicon. *J. Phys. Chem. C* **2012**, *116*, 13767-13773.

13. Peng, K. Q.; Hu, J. J.; Yan, Y. J.; Wu, Y.; Fang, H.; Xu, Y.; Lee, S. T.; Zhu, J. *Fabrication of Single-Crystalline Silicon Nanowires by Scratching a Silicon Surface with Catalytic Metal Particles*, *Adv. Funct. Mater.*, **2006**, 16, 387–39.

14. Huang, Z.; Greyer, N.; Werner, P.; de Boor, J.; Gösele, U. Metal-Assisted Chemical Etching of Silicon: A Review. *Adv. Mater.* **2011**, 23, 285-308.

15. Zhang, M.-L.; Peng, K.-Q.; Fan, X.; Jie, J.-S.; Zhang, R.-Q.; Lee, S.-T.; Wong, N.-B. Preparation of Large-Area Uniform Silicon Nanowires Arrays through Metal-Assisted Chemical Etching. *J. Phys. Chem. C* **2008**, *112*, 4444-4450.

16. Qu, Y.; Liao, L.; Li,Y.; Zhang, H.; Huang, Y.; Duan, X. Electrically Conductive and Optically Active Porous Silicon Nanowires. *Nano Lett.* **2009**, *9*, 12, 4539-4543.

17. Lin, L.; Guo, S.; Sun, X.; Feng, J.; Wang, Y. Synthesis and Photoluminescence Properties of Porous Silicon Nanowire Arrays. *Nanoscale Res. Lett.* **2011**, *5*, 1822-1828.

18. Hochbaum, A.I.; Gargas, D.; Hwang, Y.J.; Yang, P. Single Crystalline Mesoporous Silicon Nanowires. *Nano Lett.* **2009**, *9,* 10, 3550–3554.

19. Peng, K.; Lu, A.; Zhang, R.; Lee, S.-T. Motility of Metal Nanoparticles in Silicon and Induced Anisotropic Silicon Etching. *Adv. Funct. Mater.* **2008**, *18*, 3026-3035.

20. Brunauer, S.; Emmett, P. H.; Teller, E. Adsorption of Gases in Multimolecular Layers. *J. Am. Chem. Soc.*, **1938**, *60*, (2), 309–319.

21. Lowell, S.; Shields, J. E.; Thomas, M. A.; Thommes, M. *Characterization of porous solids and powders: surface area, pore size and density*, 3rd ed.; Springer: Netherland, 2006.

22. To,W.-K.; Tsang, C.-H.; Li, H.-H.; Huang, Z. Fabrication of n-Type Mesoporous Silicon Nanowires by One-Step Etching. *Nano Lett.* **2011**, *11*, 12, 5252-5258.

23. Lee, B.; Rudd, R. E. First-principles study of the Young's modulus of Si <001> nanowires, *Phys. Rev. B*, **2007**, 75, 041305(R).

24. Hoffmann, S.; Utke, I.; Moser, B.; Michler, J.; Christiansen, S. H.; Schmidt, V.; Senz, S.; Werner, P.; Gösele, U.; Ballif, C. Measurement of the Bending Strength of Vapor−Liquid−Solid Grown Silicon Nanowires, *Nano Lett.*, **2006**, Vol. 6, No. 4, 622–625.

25. Sohn, Y.-S.; Park, J.; Yoon, G.; Song, J.; Jee, S.-W.; Lee, J.-H.; Na, S.; Kwon, T.; Eom, K. Mechanical Properties of Silicon Nanowires, *Nanoscale Res. Lett.* (**2010**) 5:211–216.

26. Cheng, S. L.; Chung, C. H.; Lee, H. C. A Study of the Synthesis, Characterization and Kinetics of Vertical Silicon Nanowire Arrays on (001) Si Substrates, *J. Electrochem. Soc.*, **2008**, *155* (11), D711–D714.

27. Zhao, D.; Ying Wan, Y.; Zhou, W. *Ordered Mesoporous Materials*, Wiley-VCH Verlag GmbH: Germany, 2013.

28. Piscanec, S.; Cantoro, M.; Ferrari, A. C.; Zapien, J. A.; Lifshitz, Y.; Lee, S. T.; Hofmann, S.; Robertson, J. Raman spectroscopy of silicon nanowires, *Phys. Rev. B,* **2003,** 68, 241312(R).

29. Hull, R. *Properties of crystalline silicon*, INSPEC, The Institution of Electrical Engineers; United Kingdom, London, 1999.

30. Li, B.; Yu, D.; Zhang , S.-L. Raman spectral study of silicon nanowires , *Phys. Rev. B* **1999**, 59, 1645.

31. Peng, C.-Y.; Huang, C.-F.; Fu, Y.-C.; Yang, Y.-H.; Lai, C.-Y.; Chang, S.-T.; Liu, C. W. Comprehensive study of the Raman shifts of strained silicon and germanium, J. Appl. Phys., **105**, 083537 (**2009**); doi: 10.1063/1.3110184.

32. Süess, M. J.; Minamisawa, R. A.; Geiger, R.; Bourdelle, K. K.; Sigg, H.; Spolenak, R. Power-Dependent Raman Analysis of Highly Strained Si Nanobridges, *Nano Lett.* **2014**, 14, 1249−1254, dx.doi.org/10.1021/nl404152r .

33. Richter, H.; Wang, Z.P.; Ley, L. The one phonon Raman spectrum in microcrystalline silicon, Solid State Communications, Vol. 39, No. 5, August **1981**, 625–629, DOI: 10.1016/0038-1098(81)90337-9.

34. Campbell, I.H.; Fauchet, P.M. The effects of microcrystal size and shape on the one phonon Raman spectra of crystalline semiconductors, Solid State Communications, Volume 58, Issue 10, June **1986**, 739–741, DOI: 10.1016/0038-1098(86)90513-2.

35. Wang, R.-P.; Zhou, G.-W.; Liu, Y.-L.; Pan, S.-H.; Zhang, H.-Z.; Yu, D.-P.; Zhang, Z. Raman spectral study of silicon nanowires: High-order scattering and phonon confinement effects, *Phys. Rev. B*, **2000**, 61, 24, 16827.